\documentclass[aps,%
twocolumn,
groupedaddress]{revtex4}
\input epsf
\usepackage{epsfig}
\usepackage{amsmath}

\def\ltap{\raisebox{-.55ex}{\rlap{$\sim$}} \raisebox{.4ex}{$<$}}
\def\gtap{\raisebox{-.55ex}{\rlap{$\sim$}} \raisebox{.4ex}{$>$}}
\def\gsim{\mathrel{\gtap}}
\def\lsim{\mathrel{\ltap}}
\def\e{\mbox{e}}

\begin{document}

\title{Deflections of cosmic rays in a random component of the
Galactic magnetic field}

\author{P.G.~Tinyakov$^{a,c}$ and I.I.~Tkachev$^{b,c}$\\
$^a${\small\it Service de Physique Th\'eorique, CP 225, 
Universit\'e Libre de Bruxelles, B-1050, Brussels, Belgium }
\\ $^b${\small\it CERN Theory
Division, CH-1211 Geneva 23, Switzerland}\\ $^c${\small\it Institute for
Nuclear Research, Moscow 117312, Russia } }

\begin{abstract}
We express the mean square deflections of the ultra-high energy cosmic
rays (UHECR) caused by the random component of the Galactic magnetic
field (GMF) in terms of the GMF power spectrum. We use recent
measurements of the GMF spectra in several sky patches to estimate the
deflections quantitatively. We find that deflections due to the random
field constitute $0.03-0.3$ of the deflections which are due to the
regular component and depend on the direction on the sky. They are
small enough not to preclude the identification of UHECR sources, but
large enough to be detected in the new generation of UHECR
experiments.
\end{abstract}

\date{\today}


\maketitle

\section{Introduction}

The Galactic magnetic field (GMF) plays an important role in the
propagation of cosmic rays even at highest energies. Expected
deflections --- of order few degrees or larger --- are comparable or
exceed the angular resolution of the existing cosmic ray
experiments. Such deflections may therefore be observable. Their
understanding is crucial when searching for sources of the
highest-energy cosmic rays if the latter are charged particles.

The detailed study of deflections of ultra-high energy proton
primaries in the GMF is particularly important if the deflections in
{\em extra-galactic} magnetic fields are small. According to the
results of Refs.~\cite{Dolag} this is likely to be the case (see,
however, Ref.~\cite{Sigl}).

The Galactic magnetic field has been shown to have both regular and
turbulent components. The regular component is thought to have a
spiral structure reminiscent of the Galactic arms with one or more
reversals toward inner (and probably also outer) Galaxy and the
magnitude of order $3~\mu$G in the vicinity of the Earth
\cite{GMFmodels}. The corresponding global model of GMF was
constructed in Ref.~\cite{Stanev:1996qj} and is often used in the
discussion of propagation of cosmic ray primaries in the Galaxy.
According to this model, protons with energy $4\times 10^{19}$~eV can
be deflected in the regular GMF by $\sim 5^\circ$. There are
indications that such a coherent deflections may indeed be
present~\cite{Tinyakov:2001ir,Gorbunov:2002hk,Teshima,Tinyakov:2003nu}
in the cosmic ray data.

The random component of GMF causes the spread of arrival directions of UHECR
around the mean position, thus diluting (and potentially destroying) important
information about the actual location of the source
\cite{Kronberg:1993vk}. Under certain conditions on the magnetic field it may
also lead to the ``lensing'' of cosmic rays \cite{Harari:2002dy,Harari:2002fj}
provided the number of sources contributing to the observed UHECR flux is
small, as is favored by the statistics of clustering
\cite{Dubovsky:2000gv}. Thus, the influence of a random fields is not simply
destructive, but may give useful information on the cosmic rays and GMF
itself.

Observationally, the magnitude of the random component of GMF is
comparable to the magnitude of the regular one. However, the
deflections of cosmic ray primaries in the random field are expected
to be considerably smaller.  Indeed, if the correlation length $L_c$
of the random component is much smaller than the propagation distance
$D$, the deflections caused by the random field are proportional to
$\sqrt{DL_c}$, see e.g. Ref.~\cite{Berezinsky_book}, while the
deflections in the regular field are proportional to the distance $D$
itself. Usually the deflections in the random field are estimated as
(see e.g. Ref.~\cite{Roulet:2003rr})
\begin{equation}
\delta_r = 0.6^\circ\cdot 
\left({10^{20}~{\rm eV} \over E/Z} \right)
\left({B_r \over 4~\mu{\rm G}}\right)
\sqrt{D\over 3~{\rm kpc}}\, 
\sqrt{L_c\over 50~{\rm pc}}\; , 
\label{FstEstimeate}
\end{equation}
where $E$ is the energy of CR primary and $B_r$ is the rms value of the
random magnetic field strength. However, the global picture of GMF is
very complicated. The magnetic field may be structured on relatively
large scales, including possibility of magnetic bubbles, sheets,
filaments, magnetic ``winds'' etc., so one cannot rely on (and be
restricted to) a generic estimate uniformly over the whole sky.

The rms deflections of cosmic ray primaries in the random field are determined
unambiguously by the magnetic field power spectrum. The latter can be
extracted from observations, see e.g. Ref. \cite{Minter}. Several existing
measurements of the GMF power spectrum indicate that the correlation length
can be large in some directions on the sky. In this case the deflections are
no longer proportional to $\sqrt{DL_c}$, and the estimate (\ref{FstEstimeate})
does not apply.

The purpose of the present paper is to estimate the deflections of
UHECR primaries in the turbulent component of the Galactic magnetic
field taking into account  possible dependence of GMF parameters on the region
of the sky and the possibility of large corelation length.
We express the UHECR deflections directly through the observable
parameters characterizing the power spectrum of the magnetic field.
To this end we derive the relation between the mean square deflection
and the power spectrum (PWS) of the GMF fluctuations. The result is
most conveniently represented through the factor $R$ defined as
\begin{equation}
\frac{\delta_r}{\delta_u} = \frac{B_r}{B_{u,\bot}}\, R\; ,
\label{thR_thU_Def}
\end{equation}
where $\delta_r$ and $\delta_u$ are deflections in the random and
uniform components of GMF, respectively, and $B_{u,\bot}$ is a
projection of a uniform field onto the direction orthogonal to the
line of sight. The factor $R$ varies between 0 and 1 and is expressed
in terms of the power spectrum of the random field by
Eq.~(\ref{theta2-fin}). Moreover, we show that it is a function of a
single observable parameter $\theta_c$, the angular scale of the break
in the relevant structure function of GMF.  In the case of small
correlation length when the estimate (\ref{FstEstimeate}) applies, one
has $R \sim \sqrt{\theta_c}$. The values of $R$ in our Galaxy derived
from the existing observational data vary in the range $R\sim
0.03-0.3$. This implies typical deflections of a $4\times 10^{19}$~eV
proton in the random field of order $0.2^\circ - 1.5^\circ$ depending
on the direction.

The paper is organized as follows. In
Sect.~\ref{sect:cor.length.and.PS} we recall the relations between the
power spectrum and the correlation length and introduce the
notations. In Sec.~\ref{observations} we describe existing
observations of the random magnetic field, in particular, the
random-to-uniform ratio and the parameters of the power spectrum. In
Sect.~\ref{deflections} we turn to the deflections of UHECR in the
random magnetic field and derive the expression for the coefficient
$R$. Sect.~\ref{conclusions} summarizes our results.

\section{Turbulent field}

\subsection{Correlation length and power spectrum of GMF}
\label{sect:cor.length.and.PS}

To introduce notations and specify our assumptions, let us first
consider the statistical properties of a random magnetic field
$B_a({\bf r})$, where $a=1,2,3$.  We define the Fourier components of
the magnetic field according to
\[
B_a ({\bf r}) = 
\int d^3q B_a({\bf q}) {\rm e}^{ i {\bf rq}}. 
\]
Here $B_a$ refers to the fluctuating component of the total magnetic
field; the regular part of GMF has to be treated separately.  In what
follows we assume that the fluctuations obey Gaussian statistics and
are spatially homogeneous and isotropic.

The last two assumptions deserve a comment. The statistical
characteristics of the magnetic field are different in different sky
patches. Our analysis and results should be applied to each of these
patches separately. We assume that statistical properties of GMF are
(approximately) constant over a single patch. Also, within one patch
the magnetic field fluctuations may not be isotropic, with the
preferred direction being set by the regular component of GMF. The
present data are not sufficient to establish or rule out the isotropy
of GMF fluctuations. With the more precise data this assumption may
need to be reconsidered, and the analysis may need to be refined.

With the above assumptions, all correlators of the magnetic field can
be expressed in terms of the two-point correlation function which can
be written as
\begin{equation}
\langle B_a^{~}({\bf q})  B_b^*({\bf q}') \rangle
= { {\cal B}(q)\over 2 q^3} (\delta_{ab}-n_a n_b)\delta^3({\bf q-q'}),
\label{BB*}
\end{equation}
where $n_a = q_a/q$ is a unit vector in the direction of ${\bf q}$ and
the projection tensor ensures the divergence-free nature of the
magnetic field, $q_a B^a ({\bf q}) = 0$.  The dimensional
normalization factors are chosen in such a way that the power spectrum
${\cal B}(q)$ has physical units of $B^2$, i.e. in our case it is
measured in the units of (Gauss)$^2$. The correlation function of the
magnetic field fluctuations is defined as
\begin{equation}
\begin{aligned}
\xi(r)&=&\langle B_a({\bf r_0})B^a({\bf r_0+r}) \rangle =
\int\frac{d^3q}{q^3}\; {\cal B}(q)\; \e^{-i{\bf qr}} \hfill ~\\ &=&
4\pi\int_0^\infty\frac{dq\,}{q}{\cal B}(q)\; \frac{\sin (qr)}{qr}
~. \hfill ~~~~~~~~~~~~~~~~~~~~~
\end{aligned}
\label{variance}
\end{equation}
It determines the rms value of the field amplitude $B_r$,
\begin{equation}
B_r^2 \equiv  \langle B_a B^a \rangle = \xi(0) \; ,
\label{Br}
\end{equation}
and the correlation length $L_c$,
\begin{equation}
L_c \equiv  \frac{\int_0^\infty dr\; \xi(r)}{\xi(0)} \; .
\label{correlationlength}
\end{equation}
The energy density contained in a random component of the magnetic
field is related to the field variance as $\rho_B = B_r^2/8\pi$.

As will be discussed in Sect.~\ref{obs:pws}, in a certain range of
momenta the existing observations support a power-law behavior of
power spectra of the magnetic field fluctuations,
\begin{equation}
{\cal B}(q) \propto \frac{1}{q^{\alpha-1}}\; .
\label{Bpws1}
\end{equation}
Since the energy density in the magnetic field is finite, there has to
be a break in the pure power-law behavior, which can be parametrized as
\begin{equation}
{\cal B}(q) = \left\{
\begin{aligned}
&A\left( {q_c\over q}\right)^{\alpha_1-1} \hskip 0.5cm \mbox{at~~ $q<q_c$}\\
&A\left( {q_c\over q}\right)^{\alpha_2-1} \hskip 0.5cm  \mbox{at~~ $q>q_c$}\; ,
\end{aligned}
\right.
\label{Bpws}
\end{equation}
where $A$ is a normalization constant and $q_c$ is the momentum scale
at which the break in the spectrum occurs.  (Note that abrupt
ultraviolet and infrared cut-off can be modeled as $\alpha_2 \rightarrow
\infty$ and $\alpha_1 \rightarrow -\infty$, respectively.) 
In what follows we assume this form of GMF power spectrum.
The variance $B_r$, and consequently the energy density, converges if
$\alpha_1<1$ and $\alpha_2 >1$. Summing up contributions from both
parts of the spectrum one finds
\begin{equation}
B_r^2 = 4\pi A \; \frac{\alpha_2-\alpha_1}{(\alpha_2-1)(1-\alpha_1)} \; .
\label{Bvariance}
\end{equation}
Finiteness of the correlation length requires stronger constraint, 
$\alpha_1<0$. One then has 
\begin{equation}
L_c = {\pi\over 2 q_c}\,{(\alpha_1-1)(\alpha_2-1)\over \alpha_1\alpha_2}\; .
\label{correlationlength2}
\end{equation}
If $0<\alpha_1<1$, the correlation length diverges at small momenta
and is dominated by the largest possible distance scale in the
problem. Deflections of cosmic rays are most significant in this case.

\subsection{A physical picture and observables}
\label{sect:picture}

\begin{figure}
\epsfig{file=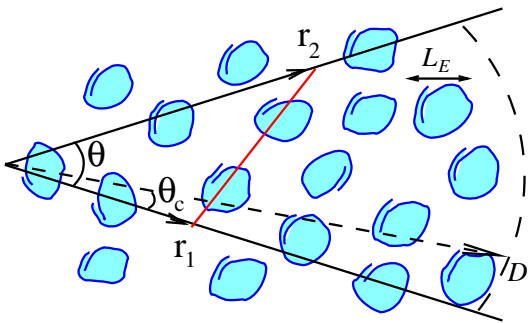,%
width=200pt,
clip=}
\caption{Schematic view of the turbulent field and the underlying geometry.
At small angles, $\theta < \theta_c = L_E/D$, there are correlated
fluctuations along the entire length of the lines of sight. At large angles
only neighboring structures contribute to correlations.}
\label{fig:scheme}  
\end{figure}

A toy model leading to the PWS with the break is illustrated in
Fig.~\ref{fig:scheme}. Here the turbulent magnetic field is contained
within the bubble-like structures. Denote by $L_E$ the typical size
of the largest energy-containing eddies, which is also called an outer
scale of turbulence or energy injection scale. It is often assumed
that $L_E$ corresponds to a typical size of the supernova remnants, and
that inside eddies the PWS corresponds to MHD or Kolmogorov
turbulence. With the assumption that fluctuations in different
``bubbles'' are uncorrelated, the corresponding correlation function
can be approximated as
\begin{equation}
\begin{array}{ll}
\xi(r) = R^{2/3}_E - r^{2/3} \quad &\mbox{at $r<R_E$},\\ 
\xi(r) = 0 & \mbox{at $r>R_E$},
\end{array}
\label{xiEddies}
\end{equation}
where $R_E = L_E/2$. Using Eq.~(\ref{correlationlength}) we see that
the correlation length of the magnetic field fluctuations is related
to the scale $L_E$ as $L_c = L_E/5$. Similar expression for the
correlation function can be also obtained from Eq.~(\ref{variance})
assuming Kolmogorov spectrum, $\alpha_2 = 5/3$, with infrared cut-off
at a momentum scale $q_c \sim 2\pi/L_E$.

In this example the 3d correlation function is zero at large
separations.  In general, however, the fluctuations may be present on
large scales as well. This is the case, e.g., when the Kolmogorov
turbulence is contained not within the ``bubbles'' as above, but
within long sheets or filaments. When substantial power is present in
fluctuations up to the largest scales, the correlation length may
diverge.

Three-dimensional power spectrum of the magnetic field fluctuations,
${\cal B}(q)$, and the corresponding 3d correlation function are not
measured directly.  Instead, one measures the parameters of the
two-dimensional angular correlation function $K(\theta)$ of some
physical observable which can be, e.g., the intensity of the polarized
synchrotron radiation or Faraday rotation measure along different
directions, see Sect.~\ref{observations} for examples. For such
observables the relation between 3d correlation function and the
angular correlation function involves integration along the line of
sight as illustrated in Fig.~\ref{fig:scheme}. For example, in the
case of Faraday rotation measure of extragalactic sources such
relation reads
\begin{equation}
K(\theta) \propto \int_0^D dr_1 dr_2 \; 
\xi_{\rm RM}^{~}(|\bf{r}_1-\bf{r}_2|).
\label{volume2angle}
\end{equation}
The directly measurable quantities are the structure
function 
\begin{equation}
\frac{1}{2} \, S(\theta) = K(0) - K(\theta) ,
\label{SF}
\end{equation}
and the two-dimensional power spectrum $C_l$ which at
small angles is related to $K(\theta )$ by the Hankel transform
\begin{equation}
K(\theta ) = \frac{1}{2\pi}\int_0^{\infty}C_l\,{\rm J_0}(l\theta )\; ldl .
\label{Cl1}
\end{equation}
Here the multipole $l$ corresponds to a typical angular scale of $\theta =
\pi/l$.

In the case of the power-law behavior of the power spectrum,
Eq.~(\ref{Bpws1}), both $S(\theta)$ and $C_l$ also follow power laws,
$S(\theta) \propto \theta^\beta$ and $C_l \propto l^{-\gamma}$. In 
a certain range of $\alpha$ (see Appendix~\ref{apendix} for
details) the three exponents are related as
\begin{eqnarray}
\label{KandC1}
&& \beta = \alpha, \\
&& \gamma =  \alpha+2 . 
\label{KandC2}
\end{eqnarray} 
Eq.~(\ref{KandC1}) is valid when $0 < \alpha < 2$. When the
correlation length is finite, $K(\theta) \rightarrow 0$ at
sufficiently large angles, and the structure function approaches
constant (cf. Eq.~(\ref{SF})). Therefore, $\beta_1 = 0$ for any
negative $\alpha_1$. On the other hand, at negative $\alpha$ the
relation (\ref{KandC2}) holds. Making use of these two relations, the
exponent $\alpha$ can be deduced from the observable quantities $\beta$
and $\gamma$. However, systematic observational effects (e.g., Faraday
depolarization and finite beam width in the case of polarized
synchrotron radiation) should be absent or accounted for.

The momentum scale $q_c$ at which the break in ${\cal B}(q)$ occurs is
not observed directly either. Instead, one observes the break in the
power-law behavior of $S(\theta)$ at some angular scale $\theta_c$ (or
the break of $C_l$ at some multipole $l_c$). The scale $q_c$ can be
estimated using the relation $l_c = Dq_c/2$, or
\begin{equation}
q_c = \frac{2\pi}{\theta_c D}\; ,
\label{qs_t}
\end{equation}
see Appendix~\ref{apendix}. These relations involve another unknown parameter,
the propagation distance in the magnetic field, $D$.  Fortunately, in the
expression for the cosmic ray deflections $q_c$ and $D$ enter as the product
$q_cD$, so the result can be conveniently expressed in terms of the observable
parameter $\theta_c$.

In the model of Fig.~\ref{fig:scheme}, the structure function at small angular
scales $\theta < \theta_c = L_E/D$ should reflect Kolmogorov turbulence,
$\beta_2 = 5/3$. On larger angular scales the structure function should
gradually become flat, $\beta_1 \rightarrow 0$. The correlation function in
this case decays on large scales as $K(\theta) \propto \theta^{-1}$, which
corresponds to $C_l \propto l^{-1}$ \cite{Cho:2002qk}.  In other words,
existence of small (compared to $D$) correlation length corresponds to
$\beta_1 = 0$ and $\gamma_1 = 1$.

\section{Observations of random component of GMF}
\label{observations} 

Current knowledge of the Galactic magnetic field is based on: (i) Faraday
rotation measurements of Galactic and extragalactic radio sources, (ii)
starlight polarization data, and (iii) observations of diffuse Galactic
synchrotron emission. Different methods are sensitive to the magnetic field in
regions with different physical conditions. Faraday rotation is sensitive to a
field in a warm ionized medium, stellar polarization measurements sample the
field in regions occupied by interstellar dust grains, while synchrotron
radiation originates from regions containing fast electrons.

Faraday rotation measure (RM) is sensitive to the projection of the
magnetic field on the line of sight. The field direction along the
line of sight is given by the sign of RM. The magnitude of the random
field $B_r$ can be estimated by analyzing the deviations of RM from
the uniform field along different directions.

Stellar and synchrotron polarization data contain information about
the field perpendicular to the line of sight. This property is
convenient for our purposes since the plane-of-the-sky component of
the magnetic field determines also the deflections of UHECR primary
particles. The ratio of the amplitudes of the random to the uniform
magnetic field components can be estimated along a single
direction. To extract the power spectrum one has to study angular
correlation function of the polarization data.

\subsection{The relative strength of uniform and random fields}
\label{BrBu}

{\it Synchrotron emission.} The total magnetic field strength is
related to the synchrotron emissivity, while the polarization of the
Galactic diffuse synchrotron background offers a method for
determining the ratio of uniform to random field strengths. Namely,
the observed fractional polarization $p_{\rm obs}$ along a given
direction obeys \cite{Ginzburg}
\begin{equation}
\frac{p_{\rm obs}}{p_{\rm max}} = 
\frac{B^2_{u,\bot}}{B^2_{u,\bot}+B^2_{r,\bot}} \; ,
\label{FracPolarization}
\end{equation}
where the subscript $\bot$ on $B_u$ and $B_r$ refers to the
plane-of-the-sky components of uniform and random field,
respectively. In Eq.~(\ref{FracPolarization}) $p_{\rm max}$ is the
fractional polarization that would be observed for a perfectly uniform
field, $p_{\rm max} \approx 0.72$. The fractional polarization of
$p_{\rm obs} \approx 35\%$ was found in Ref.~ \cite{Spoelstra} to be a
typical maximum for our Galaxy. This implies
\begin{equation}
B_{u,\bot}/B_{r,\bot} \approx 1 \; .
\label{BratioSynch}
\end{equation}
Note that this ratio depends upon direction. For instance, in the
direction of the Galactic anti-center $p_{\rm obs} \approx 20\%$ (being
averaged over $-20^\circ < b < 20^\circ$), which corresponds to
$B_{u,\bot}/B_{r,\bot} \approx 0.62$.  The typical coherence length was
estimated in Ref.~\cite{Spoelstra} to be less than $75$ pc, while the
distance to the region where polarized emission originates was found
to be about $\sim 0.5$~kpc. 

{\it Starlight polarization.} Polarization in starlight appears
because of selective absorption by interstellar dust grains whose
minor axis is aligned with the magnetic field $\bf{B}$.  The same
expression, Eq.~(\ref{FracPolarization}), is valid for the starlight
polarization data as well (with $p_{\rm max}$ being related to dust
extinction). The resulting magnitude of the random component of
magnetic field derived from the starlight polarization data is
consistent with Eq. (\ref{BratioSynch}). For example, the estimate of
Ref.~\cite{Fosalba:2001wr} reads $B_{u,\bot}/B_{r,\bot} \approx 0.8$.

{\it Faraday rotation.} Unlike polarization data, the Faraday rotation
measure is sensitive to the magnetic field component parallel to the
line of sight. Another disadvantage of this method is that it does not
allow to find the ratio of random to uniform components of the
magnetic field along a given direction. However, this information can
be extracted from the residuals of a fit to a uniform field provided
RMs in many neighboring directions are known. For instance, such kind
of study for a particular region of the sky of about $10^\circ \times
10^\circ$ centered at $(l,b) \approx (140^\circ,-40^\circ)$ was
carried out in Ref.~\cite{Minter}.  It was found that $B_r \approx
B_u$, in agreement with Eq.~(\ref{BratioSynch}).

\subsection{The power spectrum of magnetic field fluctuations}
\label{obs:pws}

{\it Synchrotron emission.}  The statistical properties of polarized
synchrotron emission depend upon direction on the sky and are
different for different observables.  In Ref.~\cite{Giardino:2002qi}
the angular power spectra (APS) of the Parkes survey of the Southern
Galactic plane at 2.4 GHz were analyzed. It was found that in the
multipole range $40<l<250$ ($0.7^\circ<\theta<5^\circ$)
the APS of $E$ and $B$ components of the polarized signal has the
slope $\gamma \approx 1.5$, and the power spectrum of polarization
angle corresponds to $\gamma \approx 1.7$. 
Similar results were found in
Refs.~\cite{Bruscoli:2002fp,Baccigalupi:2000pm} for other Galactic
latitudes.  In particular, while being close to $1.5$ on average, the
slopes of $E$ and $B$ components in the multipole range $l<1000$
($\theta>10'$) were found \cite{Bruscoli:2002fp} to be in the range $1
< \gamma < 2.7$ depending on the particular region of the sky and the
survey used.

Negative values of $\alpha$ (derived with the use of
Eq.(\ref{KandC2}), if applicable) indicate that in many sky patches
the correlation length of magnetic field may be small.

{\it Starlight polarization.}  The angular power spectrum of the
starlight polarization for the Galactic plane data ($|b| < 10^\circ$)
is consistent with $\gamma \approx 1.5$ for all angular scales $\theta
> 10'$ (or $l<1000$), see Ref.~\cite{Fosalba:2001wr}.

{\it Faraday rotation.}  Structure functions of the rotation measure
of extragalactic radio sources were studied in
Refs.~\cite{Simonetti1,Simonetti2,Minter,Clegg,Sun:2004mt}. Three
different sky patches were considered in Ref.~\cite{Simonetti1}. In
two patches the index $\beta_1$ of the structure function of rotation
measure was found to be consistent with zero on large angular scales
$> 2^\circ$, while in the third positive $\beta$ was observed. There
is a clear drop in the structure function on small angular scales
$\theta < 0.1^\circ$, \cite{Simonetti1,Simonetti2}. Therefore, the
break in the spectrum has to be at angular scales $0.1^\circ <
\theta_c < 2^\circ$ ($\theta_c$ cannot be quantified more precisely as
there are no data points at these intermediate angular scales).

In Ref.~\cite{Sun:2004mt} shallow (with $\beta_1 < 0.3$) structure
functions in several sky regions near the Galactic plain were found over the
range $0.3^\circ < \theta_c < 10^\circ$. Similar result was obtained in
Ref.~\cite{Clegg}.

Minter and Spangler \cite{Minter} have studied structure functions of
RM from polarized extragalactic sources for a particular region with
previously mapped emission measure of warm ionized medium. In this
paper fluctuations in electron density were factored out and the power
spectrum of fluctuating magnetic field was determined. The spectrum of
random magnetic field derived in Ref.~\cite{Minter} can be
parametrized by Eq.~(\ref{Bpws}) with $A \approx 4.5\times 10^{-2}\;
\mu G^2$.  At large scales the angular structure function is
consistent with the two-dimensional turbulence, $\alpha_1=2/3$, while
at small scales $q>q_c$ the spectrum coincides with the Kolmogorov
turbulence $\alpha_2=5/3$. The break in the spectrum occurs at
$\theta_c \sim 0.07^\circ$ which corresponds to $2\pi/q_c \approx
3.6$~pc assuming $D = 3$~kpc. With the parameters found in
Ref.~\cite{Minter}, Eq.~(\ref{Bvariance}) gives $B_r \approx 1.6\; \mu
G$. Note that for the uniform component of the magnetic field in the
same region one has $B_u \approx 2.2\; \mu G$ and $B_{u,||} \approx
-0.8 \; \mu G$.  The slope of $\alpha_1=2/3$ was measured up to
$2\pi/q\sim 80$~pc. Thus, in this particular sky patch the
correlation length of magnetic field fluctuations either diverges or
is larger than 80~pc.

Power spectra and the structure functions of the rotation measure
of the diffuse Galactic polarized radio background in several sky
patches near the Galactic plain were studied in
Ref.~\cite{Haverkorn:2003ad}.  Angular power spectra show a spectral
index $\gamma_1 \approx 1$, while the structure functions are
approximately flat, $\beta_1 \approx 0$ in the range $0.1^\circ <
\theta < 10^\circ$. This is indicative of the field uncorrelated on
large scales.  The structure functions may show a break at $\theta_c$
close to $0.3^\circ$, which is at the same spatial scales ($\approx
3.9$ pc assuming a path length of 600 pc) as a break in the structure
function in the RMs of extragalactic sources of Ref.~\cite{Minter}.
Note that the RM of extragalactic sources probes the complete line of
sight through the Galaxy, whereas, as a result of depolarization, the
synchrotron emission observed at low frequencies only probes the
nearby interstellar medium. In a similar study \cite{Haverkorn:2004fw}
in anther sky region, a shallow structure function of the rotation
measures was found, $\beta_1 \approx 0.2$ for $4'< \theta < 5^\circ$,
while at smaller angular scales the structure function steepens.

\section{UHECR deflections in the random magnetic field} 
\label{deflections}

In this section we show that the knowledge of the ratio
$B_{u,\bot}/B_r$, the exponents $\alpha_1$ and $\alpha_2$, and the
angular scale $\theta_c$ is sufficient to quantify the spread of
deflections of UHECR primaries caused by the random component of GMF.
As we have seen in Sect.~\ref{BrBu}, the existing observations suggest
that the magnitudes of the random and uniform components of GMF are
comparable. In what follows it will be convenient to normalize the
deflection due to the random field to the deflection $\delta_u$ which
would occur in the uniform field over the same distance and at the
same particle energy and charge.  After traveling the distance $D$ in
a uniform magnetic field, a particle with the electric charge $Ze$ and
energy $E$ is deflected by an angle \footnote{To be precise, the
product $D B_{u,\bot}$ should be replaced by the integral along the
line of sight $\int_0^D dx, B_{u,\bot(x)}$. This subtlety is
particularly important for directions close to the Galactic plane.}
\begin{equation}
\delta_u = {ZeD\over  E}\, B_{u,\bot}  \;.
\label{DeflBu}
\end{equation}
This has to be compared to the mean square deflection angle
$\delta_r$ in the random component of the Galactic magnetic filed.

\subsection{Mean square deflection in terms of magnetic field power spectrum} 

Deflections of UHECR primaries by random magnetic field were studied
in many papers, see e.g. Refs.~%
\cite{Berezinsky:1989ji,Blasi:1998xp,Sigl:1998dd,Achterberg:1999vr,%
Stanev:2000fb,Casse:2001be,Yoshiguchi:2002rb,Aloisio:2004jd}. However,
usually the main focus is the diffusive regime in the extra-galactic
magnetic field.  Deflections in the turbulent component of GMF were
studied in Refs.~\cite{Harari:2002dy,Harari:2002fj}, with the emphasis
on the possibility of magnetic field reconstruction with future high
statistics cosmic ray data. A generic turbulent component of GMF with
a simplifying assumption of a cell-like structure was also included in
the Monte-Carlo simulations of Ref.~\cite{Prouza:2003yf}. To our
knowledge, estimates of UHECR deflections in the random field based on
measurements of the MF power spectra in specific sky patches do not
exist in the literature. The UHECR deflections in the situation when
the coherence length is not small (which might be relevant for the
case of realistic GMF) was not studied in detail either.

Propagation of UHECR primaries in the Galaxy is quasi-rectilinear,
with typical deflection angles not exceeding $10^\circ-20^\circ$ even
for lowest energies. The contribution of turbulent field in these
deflections is expected to be even smaller.  Therefore, a ballistic
approximation gives a good description of UHECR propagation. In this
regime, the deflection angles are characterized by the following line
integrals,
\begin{equation}
\delta_i = {Ze\over E} \int_0^D dz \; \epsilon_{ik} B_k (z)\; ,
\label{theta}
\end{equation}
where the axis $z$ is chosen along the particle trajectory and indices
$i,k=1,2$ label two orthogonal directions. The mean square deflections are
\begin{equation}
\delta_r^2 \equiv 
\langle \delta_i\delta^i\rangle 
= {Z^2e^2\over E^2} \int\!\!\!\int_0^D dz\, dz'
\;\epsilon_{ik}\epsilon_{jp} \langle B_k (z) B_p (z')\rangle \; . 
\label{theta2}
\end{equation}
Here the average is taken over the ensemble of different realizations
of the turbulent magnetic field $B_a(x)$. For a statistically
homogeneous random field the correlator in Eq.~(\ref{theta2})
is  the function of $r = z'-z$
\begin{equation}
\epsilon_{ik}\epsilon_{jp} \langle B_k(z)B_p(z')\rangle = 
\xi_{11}(r) + \xi_{22}(r) \equiv \xi_{\bot}(r)\; ,
\label{isotropic}
\end{equation}
where $\xi_{ii}(r) \equiv \langle B_i (z) B_i (z+r)\rangle$ (no
summation over $i$).  Using Eq.~(\ref{BB*}) which enforces the
divergence-free constraint one finds
\[
\xi_{\bot}(r) = 4\pi\int_0^\infty  {dq \,  \over q} {\cal B}(q)
\left[\frac{\sin (qr)}{qr}+\frac{\cos (qr)}{q^2r^2}-\frac{\sin (qr)}{q^3r^3}
\right] \; .
\]
This relation implies 
\begin{equation}
\int_0^{\infty}dr\; \xi_{\bot}(r) = \frac{1}{2}\int_0^{\infty}dr\; \xi(r) \; .
\label{Lxibot}
\end{equation}
Note that the assumption of chaotically oriented magnetic cells, which
is often made, would give instead $\xi_{\bot}(r)= (2/3)\; \xi(r)$.
However, this assumption is inconsistent with the divergence-free
nature of the magnetic field.

Changing variables in Eq.~(\ref{theta2}) from $z$, $z'$ to $r$ and
$u=(z+z')$ one obtains
\begin{equation}
\delta_r^2 =  {2Z^2e^2\over E^2} \;B^2_r 
\int_0^{D} du \int_0^{u} dr\; \frac{\xi_{\bot}(r)}{\xi(0)}\; , 
\label{theta3}
\end{equation}
where Eq.~(\ref{Br}) was used.  It is convenient to represent the
result as a ratio of rms deflections in random field to the deflection
in the uniform field $\delta_u$ given by Eq. (\ref{DeflBu}). Thus,  
we arrive at Eq.~(\ref{thR_thU_Def}) 
where the dimensionless factor $R$ is 
\begin{equation}
R^2 \equiv \frac{2}{D^2} \;
\int_0^{D} du \int_0^{u} dr\; \frac{\xi_{\bot}(r)}{\xi(0)}\; .
\label{R_Def}
\end{equation}
This factor varies between zero  and one. 

If the correlation length $L_c$ defined by
Eq.~(\ref{correlationlength}) is much smaller than the propagation
distance $D$, the upper limit in the integral over $r$ can be extended
to infinity. One then finds 
\begin{equation}
R^2 =   \frac{L_c}{D} \; .
\label{thetaLc}
\end{equation}
In the general case, the expression Eq.~(\ref{R_Def}) can be brought to the
form
\begin{equation}
R^2 = 
\frac{4\pi}{D\xi(0)}
\int_0^\infty  {dq \,  \over q^2} {\cal B}(q) \,f(Dq)\; ,
\label{theta2-fin}
\end{equation}
where
\begin{equation}
f(x) = {\rm Si}(x) + \frac{\cos x}{x} - \frac{\sin x}{x^2}\; ,
\label{functionF}
\end{equation}
and~ ${\rm Si}(x) = \int_0^x dy\, \sin(y)/y$ ~is the integral sine
function.  At small arguments the function $f(x)$ grows linearly as
$f(x) = 2x/3 + O(x^3)$, while at $x\gsim 2\pi$ it rapidly converges to
the asymptotic value $\pi/2$.

\subsection{Mean square deflections in the random component of GMF}
\label{finalresults}

With the assumption that the power-law spectrum of the turbulent
magnetic field is given by Eq.~(\ref{Bpws}), as supported by the
existing observations, Eq.~(\ref{theta2-fin}) gives
\begin{eqnarray}
\label{Rgeneral}
&&R^2=\frac{(\alpha_2-1)(1-\alpha_1)}{(\alpha_2-\alpha_1)}\;\times\\
&&\left[(Dq_c)^{\alpha_1-1}\,\int_0^{Dq_c} {dy f(y)\over y^{1+\alpha_1}} 
+(Dq_c)^{\alpha_2-1}\,\int_{Dq_c}^\infty {dy f(y)\over y^{1+\alpha_2}} 
\right]\; ,
\nonumber
\end{eqnarray}
where $f(y)$ is defined in Eq.~(\ref{functionF}). As one can see, the final
result depends on the product $Dq_c$. Therefore, with the use of
Eq.~(\ref{qs_t}) it can be rewritten in terms of the single directly observable
scale $\theta_c= 2\pi/Dq_c$.

\begin{figure}
\epsfig{file=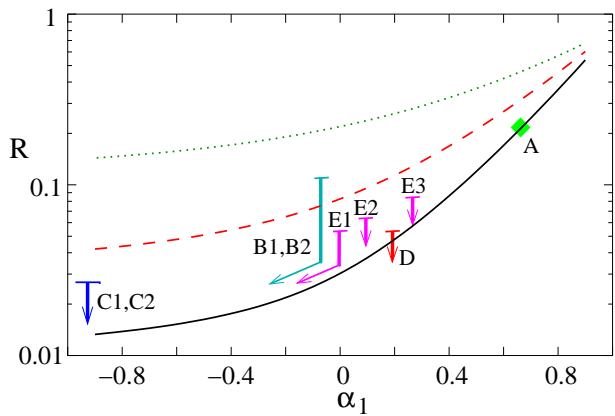,%
width=230pt,
clip=}
\caption{The coefficient R as a function of $\alpha_1$ is shown by dotted,
dashed and solid curves for $\theta_c=6^\circ, ~0.6^\circ$ and $0.06^\circ$
respectively. The data-points correspond to the APS derived for sky regions
A-E, as discussed in the text. These regions are displayed in
Fig.~\ref{fig:regions}. }
\label{fig:Ra}  
\end{figure}

In the case $Dq_c \gg \pi$ and $\alpha_1 <0$ (when both integrals are
saturated at $y=Dq_c$), we recover Eq.~(\ref{thetaLc}), 
\begin{equation}
R =\sqrt{\frac{L_c}{D}} =
\sqrt{\frac{\theta_c}{4}\, 
\frac{(\alpha_1-1)(\alpha_2-1)}{\alpha_1\alpha_2}} . 
\end{equation}
For $\alpha_1$ varying within $0.2 < \alpha_1 < 0.8$ and
$\alpha_2=5/3$ (which corresponds to the Kolmogorov turbulence), there
exists an approximate analytic expression for $R$ which holds with an
accuracy of about 10\%,
\begin{equation}
R \approx (Dq_c)^{(\alpha_1-1)/2} = (\theta_c/2\pi)^{(1-\alpha_1)/2}\; . 
\end{equation}
In the general case the factor $R$ has to be calculated numerically. Its
dependence on $\alpha_1$ for $\alpha_2=5/3$ in three cases $\theta_c=6^\circ$,
$\theta_c=0.6^\circ$ and $\theta_c=0.06^\circ$ is shown in Fig.~\ref{fig:Ra}
by the dotted, dashed and solid lines, respectively. We recall now that in
many ``low'' resolution all-sky studies the steepening of APS is not
detected, up to large multipoles, $l\sim 1000$. This suggests that $\theta_c <
10'$ and the dashed line in Fig.~\ref{fig:Ra} may serve as an  upper
limit for the factor $R$.
 
\begin{figure}
\epsfig{file=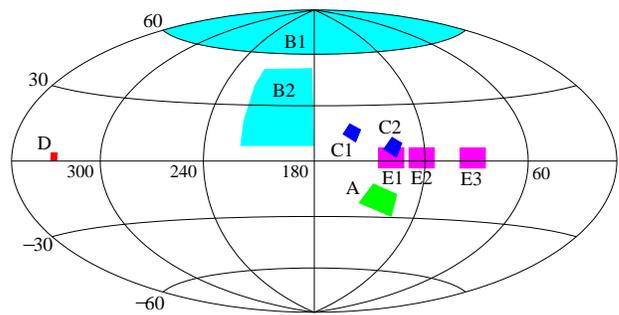,%
width=230pt,
clip=}
\caption{Regions A-E in Galactic coordinates where the break in
the APS was detected. Corresponding values of $R$ for these
regions are shown as data-points in Fig.~\ref{fig:Ra}. }
\label{fig:regions}  
\end{figure}

The labeled data-points in Fig.~\ref{fig:Ra} represent the data based
on the Faraday rotation measurements. The corresponding survey regions
are shown in Fig.~\ref{fig:regions} as colored patches labeled in the
same way as the data-points by A (Ref.~\cite{Minter}), B1,B2
(Refs.~\cite{Simonetti1,Simonetti2}), C1,C2
(Ref.~\cite{Haverkorn:2003ad}), D (Ref.~\cite{Haverkorn:2004fw}) and
E1,E2,E2 (Ref.~\cite{Sun:2004mt}). Only in the region A the transition
to the Kolgomogorov turbulence was detected at the scale $\theta_c
\sim 0.07^\circ$.  Though in some other regions an indication for the
break was observed, the transition to the spectra with $\alpha_2 > 1$
was not established. The corresponding limits on $\theta_c$ are shown
as downward arrows.  As explained in the Appendix~\ref{apendix}, any
negative $\alpha$ corresponds to $\beta = 0$ in the structure
functions. Therefore, we plot data-points with $\beta = 0$ as downward
arrows at $\alpha_1 = 0$ turning to the left (pointing to the region
of negative $\alpha_1$). Finally, in the regions C1,C2 both the
structure functions and the APS of rotation measure were obtained
\cite{Haverkorn:2003ad} resulting in $\beta \approx 0$ and $\gamma
\approx 1$.  This allows to specify $\alpha_1$ in these regions as
$\alpha_1 \lsim -1$.

Small observed values of (or upper limits on) $\theta_c$ indicate that
either the scale $2\pi/q_c$ is small, or the extent of GMF along given
direction, $D$, is large. In either case the resulting coefficient $R$
is small, $0.02 < R < 0.2$. Note that the application of
Eq.~(\ref{thR_thU_Def}) to the directions along the Galactic plane
should be done with care. Namely, deflections in the regular field
cannot be approximated by a simple relation (\ref{DeflBu}), but should
be replaced by the integral along the line of sight. Existing field
reversals can diminish deflections in the regular field considerably.

\section{Conclusions}
\label{conclusions}

Deflections of UHECR in the random component of the Galactic magnetic field
are usually discussed in the limit when the correlation length $L_c$ is much
smaller than the propagation distance $D$. However, even in this limit both
the correlation length and the propagation distance in GMF (and therefore
typical deflections) vary with direction. Moreover, the existing GMF data
suggest that the assumption of small correlation length may not be valid
uniformly all over the sky.

We have calculated deflections in a more general approach which does not
require correlation length to be small and relies directly on the spectrum of
GMF fluctuations measured in a relatively small patch of the sky.  Our method
therefore automatically takes into account variations of the GMF
characteristics with direction.  We have shown that the ratio of the
deflections in the random and uniform components of GMF,
Eq.~(\ref{thR_thU_Def}), is expressed in terms of the factor $R$ which depends
on the spectrum of the magnetic field fluctuations as given by
Eq.~(\ref{theta2-fin}). For the power-law spectrum with a single break the
factor $R$ can be written as a function of one directly observable parameter
$\theta_c$, the angular scale of the break in the relevant structure function.
In the case of a small correlation length one finds $R \sim \sqrt{\theta_c}$.

Using the measurements of the GMF power spectrum in the sky regions
where it is available, we have shown that the deflections in the
random component are small, $0.03-0.3$ of the deflections in the
uniform field, see Fig.~\ref{fig:Ra}. This is sufficiently small not
to preclude identification of sources of UHECR using methods of
Refs.~\cite{Tinyakov:2001nr,Tinyakov:2003nu,Tinyakov:2001ir}. For
instance, the deflection of a proton with energy $E=4\times 10^{19}$
eV due to the random component of GMF is expected to be about
$0.2^\circ - 1.5^\circ$ depending on the direction. This is below the
resolution of the AGASA experiment, but can be above the resolution of
the HiRes detector in the stereo mode and the expected resolution of
the Pierre Auger experiment. Thus, the detailed study of the random
component of GMF is particularly important for the interpretation of
data which will be collected by the new generation of UHECR
experiments. To this end, the all-sky map of the essential parameters
determining the power spectra of GMF is highly desirable. These maps
may be obtained from the measurements of Faraday rotation and maps of
diffuse polarized synchrotron Galactic emission.

\section*{Acknowledgments}

We are grateful to S. Dubovsky, D. Grasso, M. Haverkorn and V. Rubakov for
useful comments.  The work of P.T. is supported in part by the Swiss Science
Foundation, grant 20-67958.02 and by IISN, Belgian Science Policy (under
contract IAP V/27).
 
\appendix

\section{ 
Angular structure functions, angular power spectrum and underlying 3D power
spectrum.}
\label{apendix}

In this section we derive Eqs.~(\ref{KandC1})--(\ref{qs_t}). These 
equations are valid for a large class of observables including Faraday
rotation measures. To simplify the presentation we derive them in the case
when the correlation function is given by equation (\ref{variance}). 

The relation between 3d correlation function and angular correlation
function involves the integration along the line of sight,
\begin{equation}
K(\theta) = \int_0^D r_1^a dr_1 r_2^a dr_2 \; \xi(|\bf{r}_1-\bf{r}_2|),
\label{volume2angle1}
\end{equation}
where $\theta$ is the angle between the vectors ${\bf r_1}$ and ${\bf
r_2}$ and $a$ takes the values $0$ and $2$ in the case of point
sources and diffuse radiation, respectively. We concentrate in what
follows on the case $a=0$ relevant for all points in Fig.~\ref{fig:Ra}
except $C_{1,2}$ and $D$.

Introducing the variables $v\equiv r_1 - r_2$ and $y\equiv(r_1 +
r_2)/2$ and taking the limit of small angles one can write
Eq.~(\ref{volume2angle1}) as follows,
\begin{equation}
K(\theta) = 4\int_0^{D/2} dy \int_0^{2y} dv \;\xi(\sqrt{v^2 + y^2\theta^2}).
\label{volume2angle2}
\end{equation}
Integration over $dv$ gives
\[
K(\theta ) = 16\pi \int_0^{D/2} dy \int_0^\infty \frac{dq}{q^2}{\cal B}(q)
\]
\begin{equation}
\times
\left[\frac{\pi}{2} {\rm J_0}(yq\theta ) - \frac{\pi}{2} 
+ {\rm Si}(2yq)\right],
\label{volume2angle3}
\end{equation}
where ${\rm J_0}$ and ${\rm Si}~$ are the Bessel and integral sine functions,
respectively. According to definition (\ref{SF}), this relation gives for
the structure function $S(\theta)$:
\begin{equation}
S(\theta ) = 16\pi^2 \int_0^{D/2} dy \int_0^\infty \frac{dq}{q^2}{\cal B}(q)
\left[1- {\rm J_0}(yq\theta ) \right] .
\label{sf1}
\end{equation}
Let us assume that ${\cal B}$ is a broken power law, Eq.~(\ref{Bpws}).
Then the structure function is also a broken power law behaving as $S
\propto \theta^{\beta_1}$ and $S \propto \theta^{\beta_2}$ below and
above the angular scale $\theta_c$, respectively. The exponents
$\beta_i$ are related to exponents $\alpha_i$ in Eq.~(\ref{Bpws}) as
follows,
\begin{eqnarray}
\beta &=& 2~~~~{\rm if}~~~~\alpha > 2,\nonumber \\ 
\beta &=& \alpha~~~~{\rm if}~~~~0< \alpha < 2 ,\nonumber \\
\beta &=& \,0~~~~{\rm if}~~~~\alpha < 0 .
\label{bvsa}
\end{eqnarray}
The break occurs at 
\begin{equation}
\theta_c = \frac{2\pi}{q_c D}.
\label{qs}
\end{equation}


The angular power spectrum, $C_l$, and the angular correlation
function at small angles are related by the Hankel transform,
Eq.~(\ref{Cl1}).  If ${\cal B}$ is given by the power law (\ref{Bpws})
with $\alpha_1 < 0$ (i.e., the correlation length converges), the
integral in Eq.~(\ref{volume2angle3}) is saturated in the region $yq
\gg 1$. In this region ${\rm Si}(2yq) \rightarrow {\pi}/{2}$;
therefore, the last two terms in Eq.~(\ref{volume2angle3}) can be
neglected. (This holds whenever the correlation length is much smaller
$D$ and the integration over $dv$ in Eq.~(\ref{volume2angle2}) can be
extended to infinity.) Comparison of Eqs.~(\ref{volume2angle3})
and~(\ref{Cl1}) gives then
\begin{equation}
C^{~}_l = \frac{16\pi^3}{l^3} \int_0^{D/2} {\cal B}\left(\frac{l}{y}\right)
\, y\, dy .
\label{Cl2}
\end{equation}
Note that Eq.~(\ref{Cl2}) is generalized trivially to the case of
non-zero $a$ by replacing $y dy $ by $y^{1+2a} dy$.  For a broken power-law
behavior of ${\cal B}$, the angular power spectrum is also a broken power
law, ~$C_l \propto l^{-\gamma}$,~ where the exponents $\gamma_i$ are
\begin{equation}
\gamma_i =  \alpha_i + 2
\label{Cl3}
\end{equation}
and the break occurs at $l_c = Dq_c/2$.

\end{document}